 \documentclass[final,5p,times,twocolumn]{elsarticle} 

\usepackage{amssymb}
\usepackage{lipsum}
\usepackage{amsmath}
\usepackage{bm}
\usepackage{hyperref}

\biboptions{sort&compress}

 \journal{arXiv} 

\begin{document}

\begin{frontmatter}

\title{Nuclear mass predictions with anisotropic kernel ridge regression}

\author[FZU,PKU]{X. H. Wu\corref{cor1}}
\ead{wuxinhui@fzu.edu.cn}
\author[ANU]{C. Pan}


\cortext[cor1]{Corresponding Author}
\address[FZU]{Department of Physics, Fuzhou University, Fuzhou 350108, Fujian, China}
\address[PKU]{State Key Laboratory of Nuclear Physics and Technology, School of Physics, Peking University, Beijing 100871, China}
\address[ANU]{Department of Physics, Anhui Normal University, Wuhu 241000, China}

\begin{abstract}
  The anisotropic kernel ridge regression (AKRR) approach in nuclear mass predictions is developed by introducing the anisotropic kernel function into the kernel ridge regression (KRR) approach, without introducing new weight parameter or input in the training.
  A combination of double two-dimensional Gaussian kernel function is adopted, and the corresponding hyperparameters are optimized carefully by cross-validations.
  The anisotropic kernel shows cross-shape pattern, which highlights the correlations among the isotopes with the same proton number, and that among the isotones with the same neutron number.
  Significant improvements are achieved by the AKRR approach in both the interpolation and the extrapolation predictions of nuclear masses comparing with the original KRR approach.
\end{abstract}

\begin{keyword}
nuclear masses, anisotropic kernel ridge regression, machine-learning, kernel function
\end{keyword}

\end{frontmatter}


\section{Introduction}

Nuclear mass is one of the most fundamental properties of nucleus~\cite{Lunney2003Rev.Mod.Phys., Yamaguchi2021PPNP}, which contains a wealth of structure information~\cite{Ramirez2012Science, Wienholtz2013Nature, Roubin2017Phys.Rev.C} and also determines related nuclear reaction energies that are important in the nucleosynthesis~\cite{Mumpower2016Prog.Part.Nucl.Phys., Jiang2021Astrophys.J., Wu2022Astrophys.J.152}.
Great achievements in nuclear mass measurements have been made in recent decades, due to the development of radioactive ion beam facilities.
Up to date, the existence of more than 3300 nuclei has been confirmed~\cite{NNDC} and the masses of about 2500 among them have been measured~\cite{Wang2021Chin.Phys.C}.
However, most of neutron-rich nuclei far from the stability valley will remain beyond the experimental access in the foreseeable future.
Therefore, reliable nuclear mass predictions are desired to further understand the nuclear landscape.

Lots of efforts have been made to predict nuclear masses and to explore great unknowns of the nuclear landscape.
Global nuclear mass descriptions can be traced back to the von Weizs\"{a}cker mass formula based on the liquid drop model (LDM)~\cite{Weizsaecker1935Z.Physik}, which includes the bulk properties of nuclei quite well but lacks other effects.
Efforts have been made in extending the LDM to incorporate more effects, which are known as the macroscopic-microscopic models~\cite{Pearson1996Phys.Lett.B, Koura2005Prog.Theor.Phys., Wang2014Phys.Lett.B, Moeller2016Atom.DataNucl.DataTables},
such as the finite-range droplet model (FRDM)~\cite{Moeller2016Atom.DataNucl.DataTables} and the Weizs\"acker-Skyrme (WS) model~\cite{Wang2014Phys.Lett.B}.
The microscopic mass models have also been developed, based on the nonrelativistic \cite{Goriely2009Phys.Rev.Lett., Goriely2009Phys.Rev.Lett.a,Erler2012Nature} and relativistic density functionals~\cite{Geng2005Prog.Theor.Phys.,Afanasjev2013Phys.Lett.B,  Xia2018Atom.DataNucl.DataTables, Meng2020SCPMA,  Yang2021Phys.Rev.C, Zhang2022Atom.DataNucl.DataTables, Pan2022Phys.Rev.C}.
The root-mean-square (rms) deviations between theoretical mass models and the available experimental data~\cite{Wang2021Chin.Phys.C} reach about $0.5$~MeV in the experimentally known region, which is still not enough for accurate studies of exotic nuclear structure and astrophysical nucleosynthesis~\cite{Mumpower2016Prog.Part.Nucl.Phys.}.
In particular, for nuclei far away from the experimentally known region, the differences among the predictions of different mass models can be as large as several tens MeV.
This would lead to large uncertainties in the $r$-process studies~\cite{Mumpower2015Phys.Rev.C, Jiang2021Astrophys.J., Wu2022Astrophys.J.152, Wu2023Sci.Bull.}.

To precisely describe nuclear masses, one should in principle properly address all the underlying effects of nuclear quantum many-body systems, e.g., bulk effects, deformation effects, shell effects, odd-even effects, and even some unknown effects.
This is a great challenge for nuclear theory because of the difficulties in understanding both nuclear interaction and quantum many-body problem.
New technologies are expected to satisfy the current requirements of nuclear mass predictions.

Recently, machine learning (ML) has been widely used in physics and nuclear physics~\cite{Carleo2019Rev.Mod.Phys., Boehnlein2022Rev.Mod.Phys., He2023Sci.ChinaPhys.Mech.Astron., Zhou2024PPNP}.
Due to the special importance of nuclear mass, many ML approaches have been employed to improve its description, such as the kernel ridge regression (KRR)~\cite{Wu2020Phys.Rev.C051301, Wu2021Phys.Lett.B, Wu2024Phys.Rev.C}, the radial basis function (RBF)~\cite{Wang2011Phys.Rev.C, Niu2018Sci.Bull.}, the Bayesian neural network (BNN)~\cite{Utama2016Phys.Rev.C, Neufcourt2018Phys.Rev.C, Niu2022Phys.Rev.C}, the Gaussian process regression~\cite{Neufcourt2019Phys.Rev.Lett., Shelley2021Universe}, the principal component analysis~\cite{Wu2024SC}, etc.
Among these approaches, it is found that the KRR has the advantage that it can avoid the risk of worsening the mass predictions for nuclei at large extrapolation, due to the performance of Gaussian kernel function and ridge regression.
The KRR is a powerful machine-learning approach, which extends ridge regression to the nonlinear case by learning a function in a reproducing kernel Hilbert space~\cite{Kung2014}.
The KRR approach was firstly introduced to improve nuclear mass predictions in 2020~\cite{Wu2020Phys.Rev.C051301}.
Later on, it was extended to include the odd-even effects (KRRoe)~\cite{Wu2021Phys.Lett.B}, and had obtained the most precise machine-learning mass model at that time.
Moreover, a multi-task learning (MTL) framework, called gradient kernel ridge regression (GKRR), for nuclear masses and separation energies, was developed by introducing gradient kernel functions to the KRR approach~\cite{Wu2022Phys.Lett.B137394}.
The successful applications of the KRR approach in nuclear masses~\cite{Wu2020Phys.Rev.C051301, Wu2021Phys.Lett.B, Guo2022Symmetry, Wu2022Phys.Lett.B137394, Du2023Chin.Phys.C, Wu2024Phys.Rev.C} have also stimulated its applications in other topics of nuclear physics, including the energy density functionals~\cite{Wu2022Phys.Rev.C}, charge radii~\cite{Ma2022Chin.Phys.C, Tang2024NST} and neutron-capture cross-section~\cite{Huang2022Commun.Theor.Phys.}.

There are two major features of the KRR approach, one is the ridge regression to avoid overfitting, the other one is the kernel function.
The kernel function is important in the KRR approach because it measures the similarity between two nuclei in the nuclear mass predictions.
Beside the Gaussian kernel function, many other commonly-used kernel functions have also been applied to the KRR approach in nuclear mass predictions~\cite{Wu2023Front.Phys.}.
However, these kernel functions do not make obvious improvement comparing with the Gaussian kernel function.
The development of KRR approach in nuclear mass predictions are usually related to the further remodulation of the commonly-used kernel function.
For example, in the KRRoe approach~\cite{Wu2021Phys.Lett.B}, the kernel function is remodulated to include the odd-even effects.
In the GKRR approach~\cite{Wu2022Phys.Lett.B137394}, the kernel function is remodulated to capture the gradient information of nuclear mass surface, i.e., the separation energies.
However, all the present employed kernel functions, i.e., the original Gaussian kernel function, the kernel function with odd-even effects, and the gradient kernel function, are isotropic kernel functions.
The only reason for adopting isotropic kernel function in nuclear mass prediction is simplicity, and anisotropic kernel function has not been adopted by any research up to now.

In the present work, the anisotropic kernel ridge regression (AKRR) approach in nuclear mass predictions is developed by introducing the anisotropic kernel function into the kernel ridge regression (KRR) approach, without introducing new weight parameter or input in the training.
A combination of double two-dimensional Gaussian kernel function is adopted in the present framework.
The hyperparameters in the anisotropic kernel functions are optimized carefully by cross-validations.
The performances of the AKRR approach are compared with the original KRR approach in detail.

\section{Theoretical Framework}

In the KRR approach, the mass residual of a theoretical model, i.e., deviation between experimental and predicted mass, of the nucleus $(Z_i, N_i)$ is expressed as
\begin{equation}\label{krr_function}
  M^{\rm KRR}_{\rm res}(Z_i,N_i) = \sum_{j=1}^{m} K[(Z_i,N_i),(Z_j,N_j)]\alpha_j,
\end{equation}
where $m$ is the number of nuclei in the training set, $\alpha_j$ are weights to be determined, and $K[(Z_i,N_i),(Z_j,N_j)]$ is the kernel function.
The weights $\alpha_j$ are determined by minimizing the loss function defined as
\begin{equation}\label{krr_lose}
  L({\bm \alpha}) = \sum_{i=1}^{m} \left[M^{\rm KRR}_{\rm res}(Z_i,N_i) - M^{\rm Data}_{\rm res}(Z_i,N_i)\right]^2 + \lambda||{\bm \alpha}||^2,
\end{equation}
which yields
\begin{equation}\label{krr_weight}
  {\bm \alpha} = ({\bm K}+\lambda{\bm I})^{-1}{\bm M}^{\rm Data}_{\rm res}.
\end{equation}
The $\lambda$ in loss function~\eqref{krr_lose} is a hyperparameter that determines the regularization strength.

The kernel function adopted in the original KRR approach~\cite{Wu2020Phys.Rev.C051301} is the isotropic Gaussian kernel function defined as
\begin{equation}\label{krr_Gkernel}
  K[(Z_i,N_i),(Z_j,N_j)] = \exp\left[ -\frac{(Z_i-Z_j)^2+(N_i-N_j)^2}{2\sigma^2} \right],
\end{equation}
where the $\sigma$ is a hyperparameter defining the length scale on the distance that the Gaussian kernel affects on the nuclear chart.
A direct extension for Eq.~\eqref{krr_Gkernel} is to introduce both $\sigma_N$ and $\sigma_Z$ in the kernel function to distinguish the length scales on the distance in the neutron direction and the proton direction, respectively, and the kernel function is remodulated as a single two-dimensional Gaussian kernel function
\begin{equation}\label{Akrr_Gkernel}
  K[(Z_i,N_i),(Z_j,N_j)] = \exp\left\{ - \left[\frac{(N_i-N_j)^2}{2\sigma_N^2} + \frac{(Z_i-Z_j)^2}{2\sigma_Z^2} \right] \right\}.
\end{equation}
If $\sigma_N\neq \sigma_Z$, this kernel would have different affecting length scales in the neutron direction and the proton direction.
Since the neutron and proton directions on the nuclear chart are naturally believed to have similar impacts, a complementary term of the kernel function \eqref{Akrr_Gkernel} is introduced to keep the same affecting length scale in the neutron direction and the proton direction.
The kernel function is remodulated as a combination of double two-dimensional Gaussian kernel function
\begin{align}
  K[(Z_i,N_i),(Z_j,N_j)] = & \exp\left\{ - \left[\frac{(N_i-N_j)^2}{2\sigma_1^2} + \frac{(Z_i-Z_j)^2}{2\sigma_2^2} \right] \right\} \notag \\
  + & \exp\left\{ - \left[\frac{(N_i-N_j)^2}{2\sigma_2^2} + \frac{(Z_i-Z_j)^2}{2\sigma_1^2} \right] \right\}. \label{Akrr_GDkernel}
\end{align}
This kernel function has the same affecting length scale in the neutron direction and the proton direction, but has different length scales in the tilted directions on the nuclear chart.

In principle, other $\sigma$'s, i.e., those not equal to $\sigma_1$ or $\sigma_2$, can be introduced in the second term of Eq.~\eqref{Akrr_GDkernel} to build a more sophisticated anisotropic kernel function.
This treatment would bring two additional hyperparameters, making the model more complicated and causing difficulty in optimizing the hyperparameters.
Therefore, it is not considered in this work.

\section{Numerical Details}

In the present work, the experimental masses are taken from the AME2020~\cite{Wang2021Chin.Phys.C}, and only the nuclei with experimental uncertainty $<100$~keV are considered.
The theoretical masses are taken from the WS4 mass model~\cite{Wang2014Phys.Lett.B}.
The total dataset includes 2340 nuclei with $Z\geq8$ and $N\geq8$.

The hyperparameter $\lambda$ has been determined in the original KRR approach~\cite{Wu2020Phys.Rev.C051301}, whose value is $0.3$.
In the AKRR approach, the same value of $\lambda$ is adopted for a better comparison.
Other hyperparameters, i.e., $\sigma$ for the KRR approach and $(\sigma_1,\sigma_2)$ for the AKRR approach, would be determined by the leave-one-out cross validation.
With a given set of hyperparameters, the refined WS4 results with the KRR (AKRR) correction for each of the 2340 nuclei can be obtained with the KRR (AKRR) network trained on all other 2339 nuclei and the rms deviation can be calculated.
The optimized set of hyperparameters is then determined according to the minima of the rms deviations.

\section{Results and discussion}

\begin{figure}[!h]
\centering
\includegraphics[width=\linewidth]{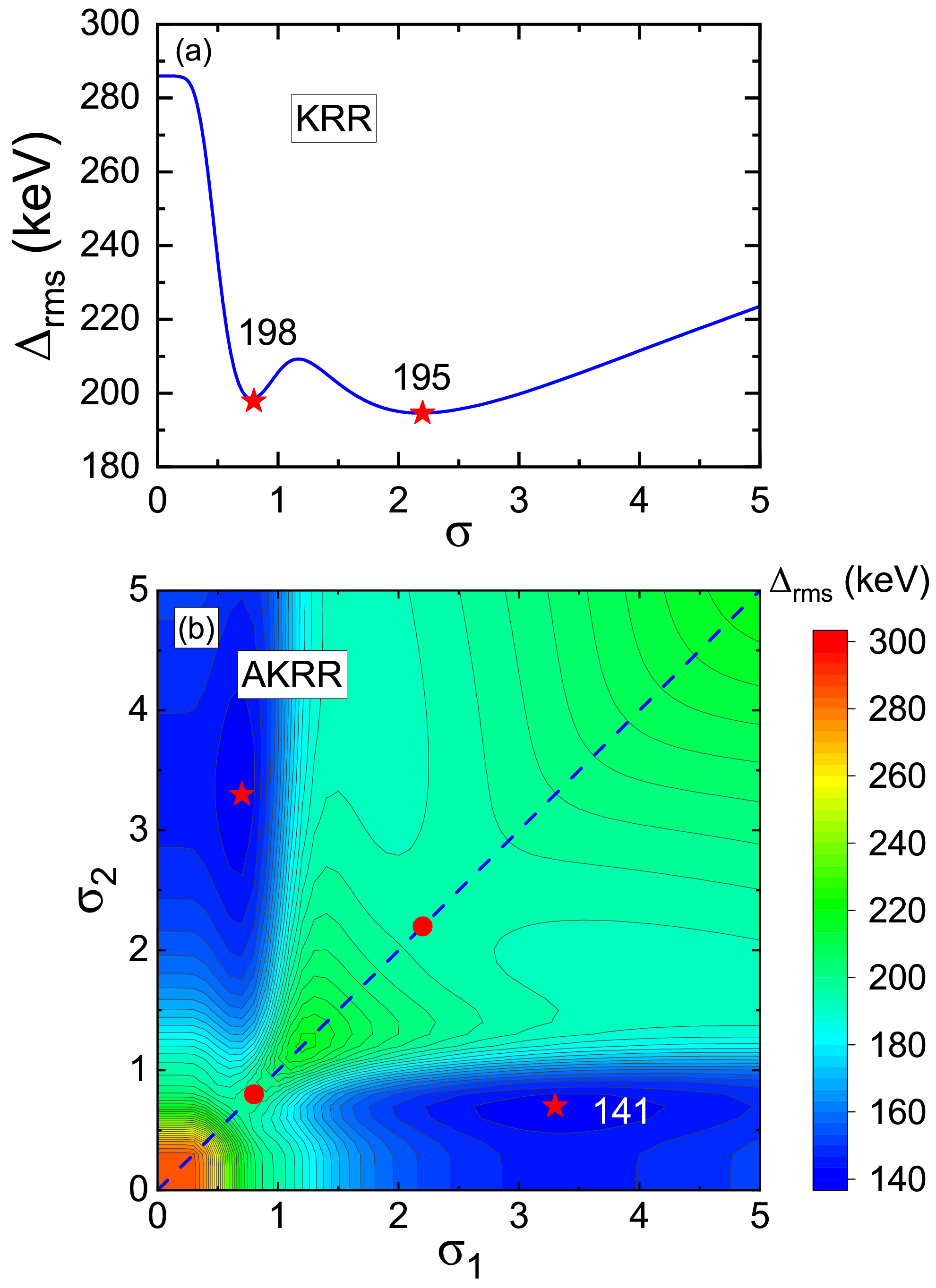}
\caption{
The rms deviations $\Delta_{\rm rms}$ obtained by the leave-one-out cross-validation with different hyperparameters.
(a) KRR approach,
(b) AKRR approach with double Gaussian kernel function.
The minima are marked by red star symbols.
The dashed line in (b) signs the cases with $\sigma_1=\sigma_2$, with which the AKRR approach would reduce to the KRR approach.
The filled circles in (b) sign the cases with $\sigma_1=\sigma_2=0.8$ and $\sigma_1=\sigma_2=2.2$.
The values signed around the minima represent the corresponding $\Delta_{\rm rms}$ respectively.
}
\label{fig1}
\end{figure}

The rms deviations obtained by the KRR approach and the AKRR approach in the leave-one-out cross-validation with different hyperparameters are shown in Fig.~\ref{fig1}.
For the KRR approach, there are two minima in the $\Delta_{\rm rms}$ curve, which is the same as in Fig.4 of Ref.~\cite{Wu2020Phys.Rev.C051301}.
These two minima correspond to a narrower ($\sigma=0.8$) Gaussian kernel and a wider ($\sigma=2.2$) one respectively.
The accuracies achieved by the KRR approach with these two minima are similar, which are around $195$~keV.

For the present AKRR approach, the $\Delta_{\rm rms}$ surface are symmetric along the line with $\sigma_1=\sigma_2$, with which the AKRR approach reduces to the KRR approach.
As can be seen in Fig.~\ref{fig1} (b), the cases with $\sigma_1=\sigma_2$ actually contribute to local maxima in the $\Delta_{\rm rms}$ surface, and the minima obtained by the KRR approach, i.e., $\sigma_1=\sigma_2=0.8$ and $\sigma_1=\sigma_2=2.2$, are just saddle points in the $\Delta_{\rm rms}$ surface.
The global minimum in the $\Delta_{\rm rms}$ surface by the AKRR approach is located at $(\sigma_1=3.3, \sigma_2=0.7)$, or equivalently, at $(\sigma_2=0.7, \sigma_1=3.3)$, achieving the accuracy of $141$~keV, which is significantly lower than the $195$~keV achieved by the KRR approach.
This shows that the breaking of the isotropic symmetry of the kernel function can help to achieve better accuracy in nuclear mass predictions, indicating the importance of introducing anisotropic kernel function in the AKRR framework.

\begin{figure}[!h]
\centering
\includegraphics[width=8.5cm]{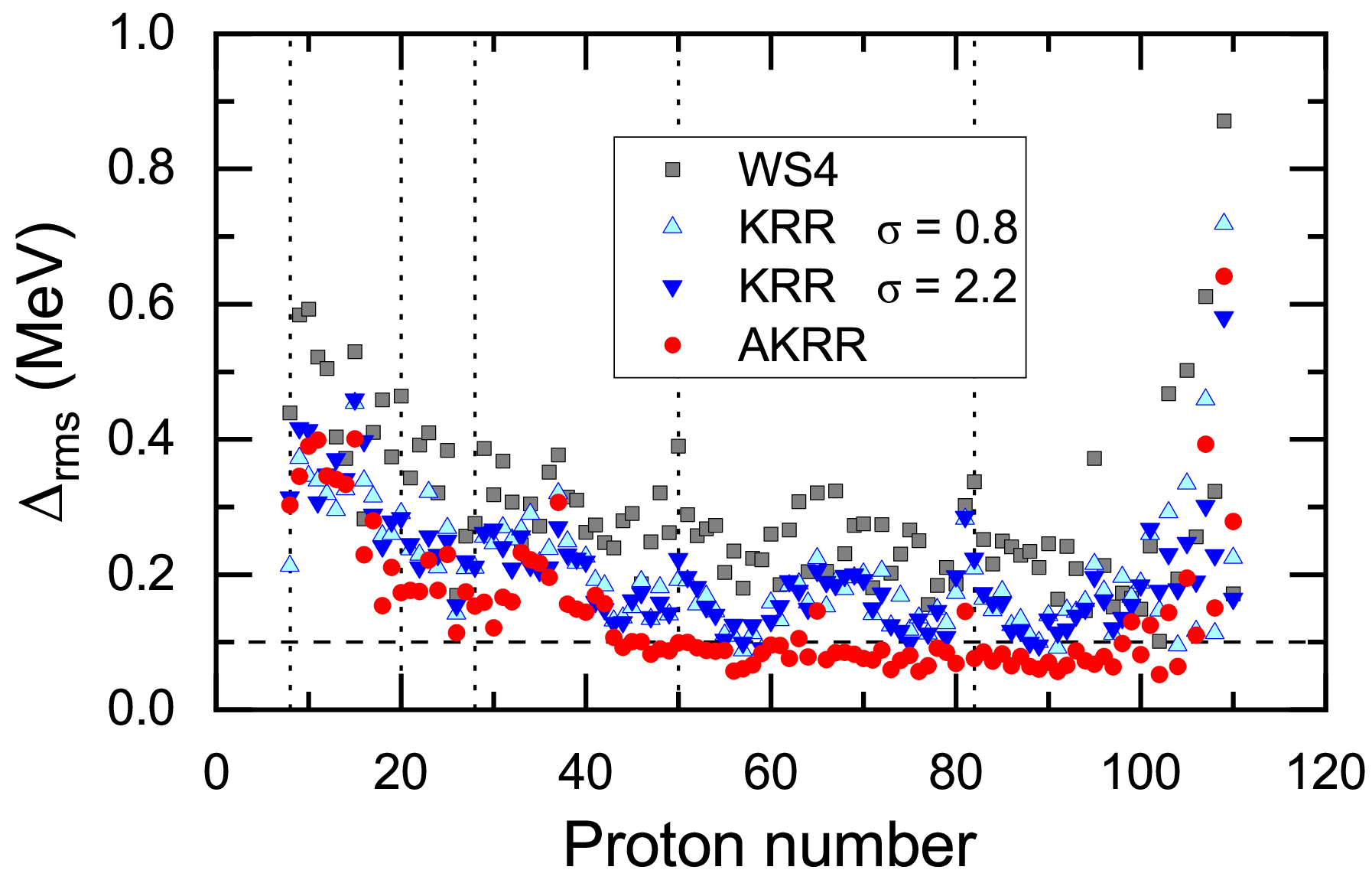}
\caption{
The rms deviations $\Delta_{\rm rms}$ for each isotopic chain between the experimental mass data~\cite{Wang2021Chin.Phys.C} and the predictions of the WS4, KRR, and AKRR approaches.
The magic numbers are indicated by vertical dotted lines.
The horizontal dashed line is drawn for guiding eyes at the rms deviation of 0.1 MeV.
}
\label{fig2}
\end{figure}

The rms deviations $\Delta_{\rm rms}$ for each isotopic chain between the experimental mass data~\cite{Wang2021Chin.Phys.C} and the predictions of the WS4, KRR, and AKRR approaches are shown in Fig.~\ref{fig2}.
Generally speaking, the deviations obtained by these approaches are smaller when coming to heavier isotopes, which is excepted for the isotopic chains with $Z>100$, where only a few experimentally known mass data are available and the $\Delta_{\text{rms}}$ could be misled by several nuclei with large deviations.
The differences between the experimental data and the WS4 mass model are roughly $0.3$~MeV for most nuclei.
The KRR approach improves the WS4 predictions, and the corresponding differences are mainly reduced to around $0.2$~MeV.
Note that in the leave-one-out cross-validation, the KRR approach with different widths of Gaussian kernel lead to similar performances.
The mass descriptions in all isotopic chains can be further improved by the AKRR approach, in particular, the differences for the medium-mass and heavy nuclei are mostly reduced to within $0.1$~MeV.

\begin{figure}[!h]
\centering
\includegraphics[width=8.5cm]{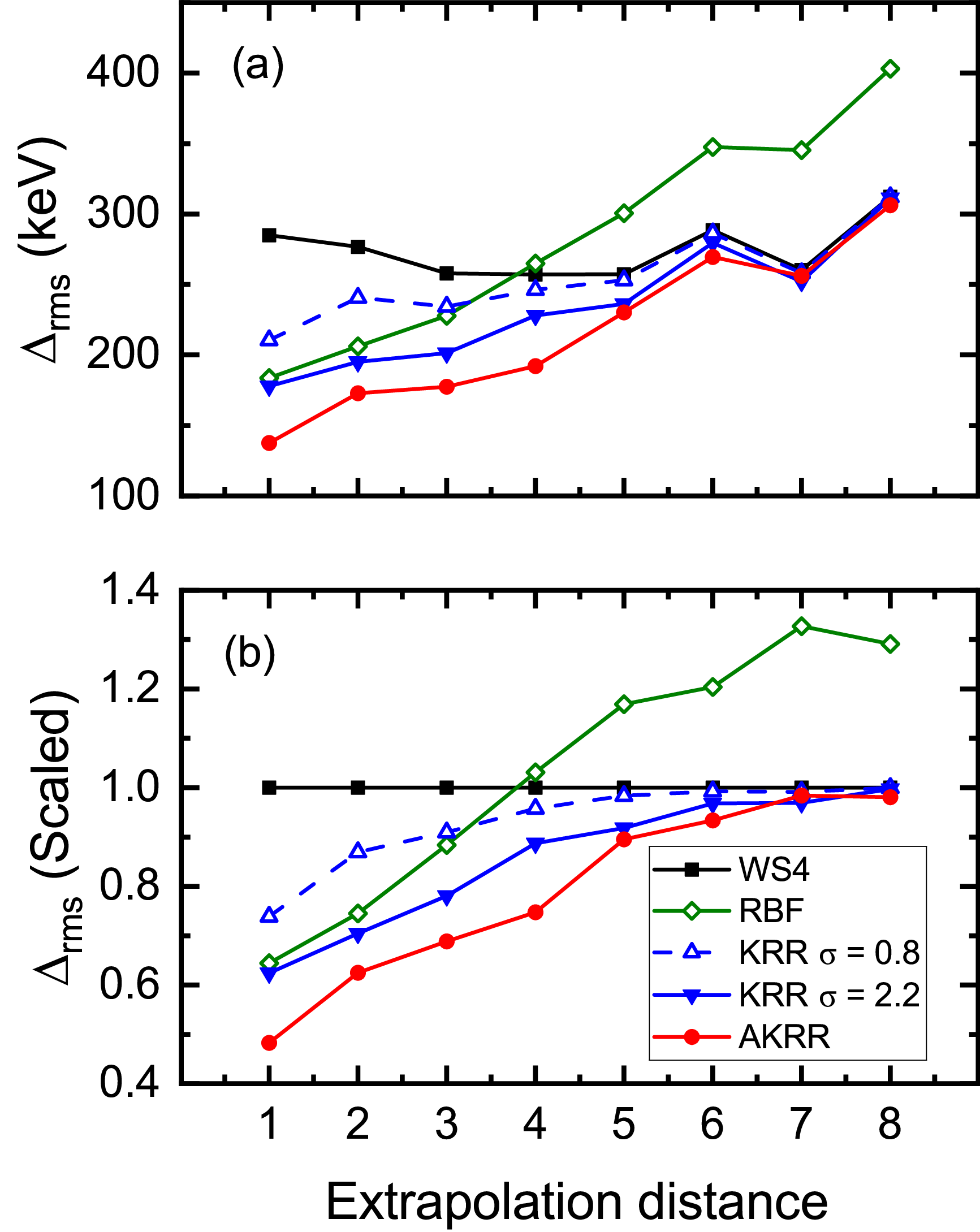}
\caption{
Comparison of the extrapolation power of the RBF, KRR, and AKRR approaches for eight test sets with different extrapolation distances (see text for details).
(a) The rms deviations $\Delta_{\rm rms}$ of the calculated masses from the WS4 mass model (filled squares), the RBF (empty diamonds), the KRR with $\sigma=0.8$ (empty triangles), the KRR with $\sigma=2.2$ (filled triangles), and the AKRR (filled circle) extrapolations with respect to the experimental masses.
(b) For a better comparison, the rms deviations $\Delta_{\rm rms}$ are scaled to the values for the WS4 mass model.
}
\label{fig3}
\end{figure}

To examine the extrapolation power of the AKRR approach for neutron-rich nuclei, similar to the Fig.3 of Ref.~\cite{Wu2020Phys.Rev.C051301}, for each isotopic chain, the eight most neutron-rich nuclei are removed from the training set, and they are classified into eight test sets respectively, corresponding to the different extrapolation distances from the remain training set in the neutron direction.
This can examine the extrapolation abilities of machine-learning approaches to neutron-rich side, which is important for the $r$-process studies~\cite{Kajino2019Prog.Part.Nucl.Phys., Cowan2021Rev.Mod.Phys.}.

In Fig.~\ref{fig3}(a), the rms deviations $\Delta_{\rm rms}$ of the calculated masses for the eight test sets from the WS4 mass model, the AKRR extrapolations, the KRR ones, and the RBF ones with respect to the experimental masses are shown as functions of the extrapolation distance.
An even more clear comparison is shown in Fig.~\ref{fig3}(b), where the rms deviations $\Delta_{\rm rms}$ are scaled to the corresponding ones for the WS4 mass model.
One can see that all the four approaches improve the mass descriptions of nuclei with smaller extrapolation distances, i.e, smaller than $4$.
When extrapolating to large distances, distinct features appear between the RBF approach and the KRR-based approaches.
The rms deviations of the RBF extrapolations are larger than the ones for the WS4 mass model, which means the RBF appraoch actually worsens the WS4 masses, while the KRR-based approach can avoid the risk of worsening the mass description for nuclei at large extrapolation distances.
This behavior has been discussed in detail in Ref.~\cite{Wu2020Phys.Rev.C051301} for the KRR approach, and it remains for the AKRR approach.

The KRR approach with $\sigma=0.8$ has indistinctive improvements, especially for the nuclei with larger extrapolation distance, which is limited by the short affecting range of the narrow Gaussian kernel.
With the proper width of Gaussian kernel $\sigma=2.2$, the KRR approach can achieve better accuracies in the extrapolation validation.
It is surprising to find that the AKRR performs globally better than the KRR approach, which indicates the effectiveness of introducing anisotropic kernel in the nuclear mass predictions.

\begin{figure}[!h]
\centering
\includegraphics[width=8.5cm]{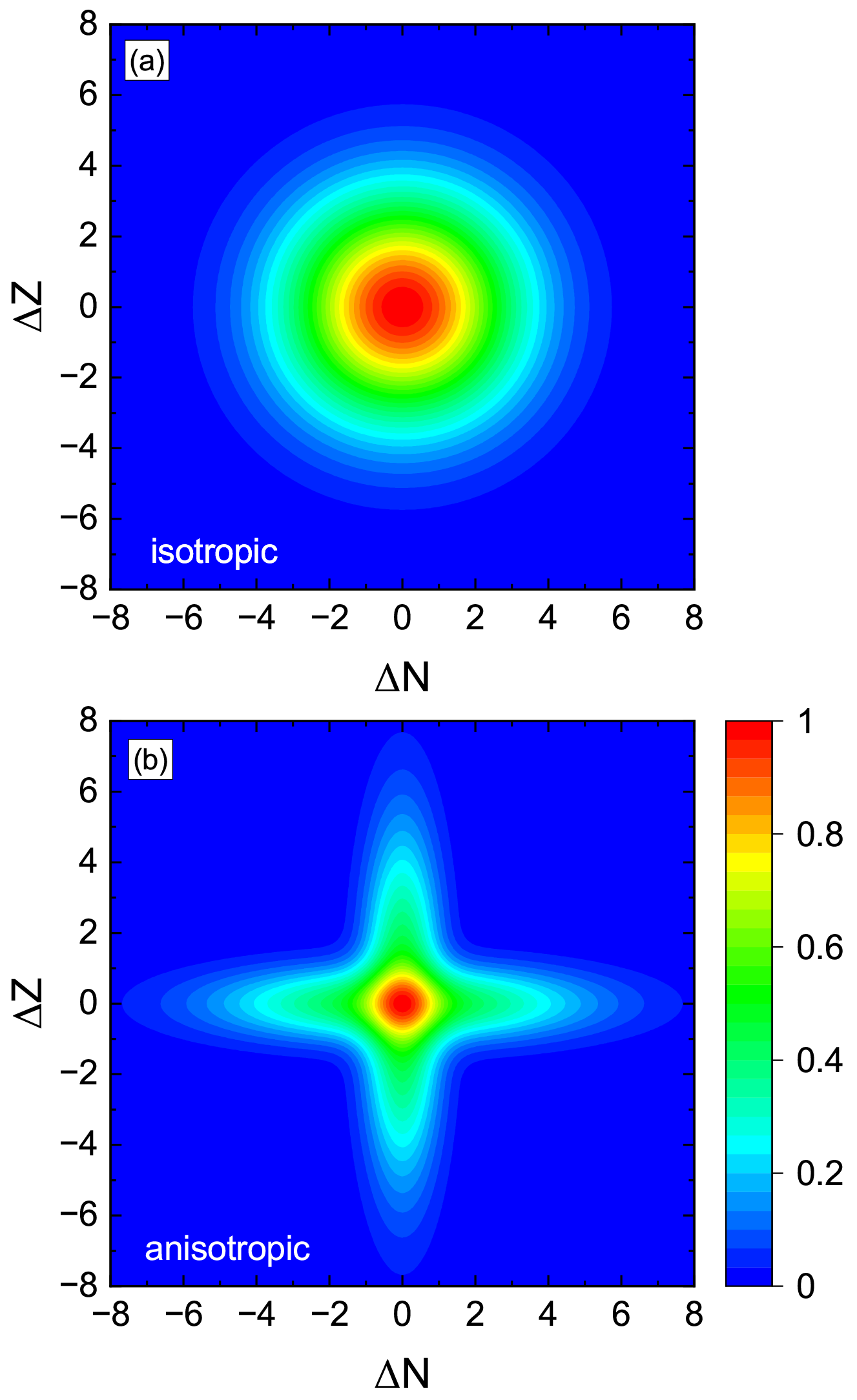}
\caption{
Distributions of the kernel functions.
(a) isotropic Gaussian kernel function with $\sigma=2.2$ adopted in the KRR approach.
(b) anisotropic double Gaussian kernel function with $(\sigma_1=3.3,\sigma_2=0.7)$ adopted in the AKRR approach.
}
\label{fig4}
\end{figure}

In the KRR framework, the kernel function measures the correlation between every two nuclei in the nuclear chart, or in other words, it quantifies the impacts of each nucleus on other ones.
Figure~\ref{fig4} shows the distributions of the isotropic Gaussian kernel function~\eqref{krr_Gkernel} with $\sigma=2.2$ adopted in the KRR approach and the anisotropic double Gaussian kernel function~\eqref{Akrr_GDkernel} with $(\sigma_1=3.3,\sigma_2=0.7)$ adopted in the AKRR approach.
As can be seen in both Fig.~\ref{fig4}(a) and Fig.~\ref{fig4}(b), the major impacts of both kernel functions are located in the region with $|\Delta N|<6$ and $|\Delta Z|<6$, which means that the mass of a certain nucleus $(N,Z)$ has little reference value for the nuclei with distances more than $6$ in the nuclear chart.
This can help to understand the extrapolation behaviors that have been discussed in Fig.~\ref{fig3}.

Distinct features of the isotropic and anisotropic kernel distributions can be clearly seen in the Fig.~\ref{fig4}(a) and Fig.~\ref{fig4}(b).
The isotropic kernel has the same impacts in all directions, while the anisotropic kernel shows cross-shape pattern, which has larger impacts in the directions with $\Delta N = 0$ and $\Delta Z = 0$, and has smaller impacts in the tilted directions on the nuclear chart.
The reason why the AKRR approach can achieve better accuracy in nuclear mass predictions can be now attributed to the anisotropic kernel function, which enhances the impacts in the the directions with the same neutron or proton numbers.

Up to this point, it is clear that the AKRR results exhibit better performances compared with the original KRR.
Here comes the question: why anisotropic kernel is better than the isotropic one?
Here are two remarks:
\begin{enumerate}
  \item[(i)] First, from the view of technicality, the anisotropic kernel is an extended version of the isotropic one, and the hyperparameter space of the anisotropic kernel is larger than the isotropic one.
             Therefore, the AKRR approach naturally has the potential to further improve the mass predictions, if the hyperparameters are finely tuned.
  \item[(ii)] Second, from the view of physics, the cross-shape pattern of the kernel function in Fig.~\ref{fig4}(b) highlights the correlation among the isotopes with the same proton number, and that among the isotones with the same neutron number.
      This correlation seems to be nature, as one typically expects a nucleus to be more similar with its isotopes or isotones than other ones.
      For example, a nucleus $(Z,N)$ should be more similar with $(Z+2,N)$ or $(Z,N+2)$ than $(Z+1,N+1)$, despite the latter having a shorter Euclidean distance in the nuclear chart.
      This similarity could be simply understood in a mean-field view, where the different isotopes (isotones) share the same or close structure of the single-proton (neutron) levels, i.e., shell structure.
      Specifically, in an isotopic chain with a magic proton number, most nuclei share the features such as the shape is spherical or near-spherical, the proton separation energy is large, etc., and the similar for an isotonic chain.
      In the AKRR approach, by strengthening the correlation among nuclei with the same neutron number or those with the same proton number, more information on the shell effect is incorporated.
      This can actually be corroborated by the features shown in Fig.~\ref{fig2}.
      The deviations given by the WS4 and KRR approaches show patterns correlated to the shell structure, while this behavior is eased by the AKRR approach.
      It should also be mentioned that in the region of light nuclei, with the increase of nucleon number the evolution of single-nucleon structure is faster than that in the region of heavier nuclei, the similarity between a nuclide with its isotopes or with its isotones is lower, and the corresponding correlation should be weaker.
      This leads to that the improvement of AKRR for light nuclei is not as significant as that for heavier nuclei, as shown in Fig.~\ref{fig2}.
\end{enumerate}

\section{Summary}

In summary, the anisotropic kernel ridge regression (AKRR) approach in nuclear mass predictions is developed by introducing the anisotropic kernel function into the kernel ridge regression (KRR) approach, without introducing new weight parameter or input in the training.
A combination of double two-dimensional Gaussian kernel function is adopted, and the corresponding hyperparameters are optimized carefully by cross-validations.
The anisotropic kernel shows cross-shape pattern, which strengthens the correlations among nuclei with the same neutron number or those with the same proton number, more information on the shell effect is incorporated.
It is found that the mass descriptions in the whole range of nuclear chart can be further improved by the AKRR approach, and the resulting rms mass deviation from the experimental data is reduced from $195$~keV for the KRR approach to $141$~keV for the AKRR approach.
Especially, the mass deviations of the medium-mass and heavy nuclei are mostly reduced to within 0.1 MeV.
It is also found that the AKRR performs globally better than the KRR approach in the extrapolation validations.
These results indicate the effectiveness of introducing anisotropic kernel in the nuclear mass predictions, which can also stimulate the applications in other nuclear properties.


\section*{Acknowledgments}

This work was partly supported by the State Key Laboratory of Nuclear Physics and Technology, Peking University under Grant No. NPT2023KFY02, the China Postdoctoral Science Foundation under Grant No. 2021M700256, and the start-up Grant No. XRC-23103 of Fuzhou University.

\bibliographystyle{elsarticle-num}
\bibliography{paper}

\end{document}